# Systematic Builder for All-Atom Simulations of Plastically Bonded Explosives

Chunyu Li[1],[*] Brenden W Hamilton[1],[*] Tongtong Shen[1], Lorena Alzate[2], Alejandro Strachan[1] [†]

1: School of Materials Engineering and Birck Nanotechnology Center, Purdue University, West Lafayette, Indiana, 47907 USA
2: Computational Science and Engineering Division, Oak Ridge National Laboratory, Oak Ridge, TN.

## Abstract

The shock to detonation transition in heterogeneous plastically bonded explosives is dominated by energy localization into hotspots that arise from the interaction of the shockwave with microstructural features and defects. The complex polycrystalline structure of these materials leads to a network of hotspot that can coalesce into deflagration and detonation waves. Significant progress has been made on the formation and potency of hotspots using atomistic simulations, but most of the work has focused on ideal and isolated defects. Hence, developed a method, denoted PBXGen, to build realistic PBX microstructures for all-atom simulations. PBXGen is generally applicable, and we demonstrate it with two systems: an RDX-polystyrene PBX with a 3D microstructure and a TATB-polystyrene with columnar grains. The resulting structure exhibit key features of PBXs, albeit at smaller scales, and are validated against experimental mechanical and shock properties.

[*] Equal contributions
[†] Email: strachan@purdue.edu

# 1. Introduction

The shock compression of materials can drive a variety of events such as phase transformations [1–3], plasticity [4–7], melting [8,9], intra-molecular deformations [10,11], and chemical reactions [12–15]. Most of these phenomena can be exacerbated by the localization of energy due to the interaction of the shockwaves with the materials microstructure and defects. A key example of this is the formation of hotspots in composite energetic materials, which accelerates chemical reactions and controls initiation and sensitivity [6,16–20]. In the shock initiation of energetic materials, the bulk shock temperature is typically not enough to prompt chemistry on fast enough timescales to cause a run to detonation. Hotspots allows for more prompt reactions in small regions that can grow and coalesce into a deflagration wave and potentially a detonation [18]. Plastic bonded explosive (PBX) microstructures are often plagued with voids, both between grains and within them, numerous cracks, grain boundaries, and various other defects that will lead to complex hotspots upon shock compression [21,22]. Several mechanisms are known to localize energy as shock induced hotspots [23], including the collapse of porosity, shear band formation, crack formation, friction, jetting, and viscous flow heating [5,23–25].

The collapse of porosity was found to be one of the dominate mechanisms for hotspot formation through shock desensitization experiments in which initial, weak compaction waves remove a significant fraction of the porosity, rendering the sample non-detonable from subsequent shocks [26]. Comparisons of inclusions (silica microbeads) and pores (air bubbles) in gelled nitromethane have also shown the superiority of the latter in forming critical hotspots. In addition, these studies have shown that more numerous, smaller pores lead to shorter run to detonations than fewer, large pores [17]. Despite this progress, the relative potency of the various mechanisms above are not known, nor are the possible interactions between them. For example, localized plastic deformation or interfacial shear can weaken shocks and weaken the effect of porosity collapse. At the same time, the combination of neighboring hotspots caused by different mechanisms can result in larger and more reactive hotspots.

The spatial and temporal scales of hotspots make their experimental characterization extremely challenging, but progress is occurring. For example, recent experimental work from Bassett et al. have measured the peak temperatures of hotspots formed via shock compression, measuring values that are sensitive to microstructure and in the range 4000-7000 K [27,28]. However, these techniques currently can only measure the peak temperature values and are unable to directly extract the mechanisms responsible for hotspot formation.

Given the experimental challenges, computational modeling of hotspot formation has attracted significant interest and provided much of our current mechanistic understanding. Hydrodynamic calculations provided a first picture of the complexity of pore collapse [29–32]. With increasing computational power and due to its explicit simulation of chemical reactions and thermo-mechanical processes with atomic detail, reactive molecular dynamics (MD) has been a key player in understanding and predicting the shock-induced reactivity [33–38]. For example, MD simulations revealed the importance of jetting and recompression as a mechanism of energy localization during pore collapse [25] and the importance of pore size and shape [39]. However, the large majority of MD simulations of hotspot formation and criticality have been limited to simplified geometries, such as single crystals with cylindrical and diamond shaped pores, due to computation limitations [39–44]. To bridge the gap between atomistic simulations and experimental measurements, continuum and mesoscale work have also been applied to model hotspot formation mechanisms such as pore collapse, crack propagation, and friction [24,45–49].

These simulations have been used to characterize simple isolated defects or microstructural features [50–55] and also realistic microstructures [56–60]. However, these methods necessarily approximate many of the underlying mechanisms responsible for energy localization, including localized plastic deformation, amorphization, jetting, pore collapse, and recompression. Overall, numerous open questions remain regarding the formation, interaction, and criticality of shock induced hotspots.

All atom simulations of realistic microstructures can be the key to proving definite answers to these longstanding issues of hotspot formation in shocked PBXs. While current computational capabilities restrict MD simulations to length scales below those in most PBX systems of interest, we believe scaled down atomistic PBX models will provide invaluable insight to the relative importance of different hotspot mechanisms. In this paper we introduce PBXGen, a PBX microstructure builder for all-atom calculations, characterize the resulting microstructures for various composite HE materials and use them with MD simulations to characterize their thermo-mechanical response. The main steps of the PBX builder are: coarse grain granular packing, HE grain preparation, polymer coating of HE grains, and insertion of polymer-coated HE grains and system equilibration and compaction. The resulting structures have densities and microstructural features comparable to those of actual PBXs. Molecular dynamics simulations of the mechanical properties and hotspot formation in these structures reveal the importance of inter-grain porosity with high aspect ratios to allow for significant molecular jetting.

## 2. Simulation Methods

PBXGen is implemented by utilizing the LAMMPS [61] simulation package for a majority of steps, and all subsequent MD simulations are performed with LAMMPS. All energetic materials and polymer binders are described with non-reactive, valence force fields that describe interactions in terms of covalent, van der Waals and electrostatic energies. Specifically, polystyrene (PS) is modeled with the Dreiding potential [62], TATB is modeled with the potential from Bedrov et. al. [63–65], and RDX is modeled with the Smith-Bharadwaj potential [66] with the modifications from Ref [39]. The cross pairs for the van der Walls interactions spanning two different force fields are added following the Dreiding potential rules. The atomic charges for PS are obtained using the Gasteiger method [67] as implemented in Polymer Modeler [68]. Long range electrostatic interactions are calculated by the particle-particle particle-mesh (PPPM) method, which is an accurate and computationally efficient method developed by Hockney and Eastwood [69,70]. A timestep of 1.0 fs is used for RDX based systems and a timestep of 0.2 fs for TATB based systems. All boundary conditions periodic except in the case of free surfaces, which utilize shrink-wrapped boundaries.

## 3. PBX Builder: PBXGen

Typically, PBXs consist of 90-95% HE particles, such as RDX, HMX, or TATB, and 5-10% polymer binder. Crystalline HE particles are usually on the scale of micrometers to millimeters, with a bimodal size distribution [22,71]. Polymer binders with a relatively lower average molecular weight, 2500-3000 g/mol, are preferred for processing [72]. The manufacturing process of a PBX has several key steps including mixing HE in powder form with a polymeric resin, curing the mixture at controlled temperature, and casting into a desired shape under pressure [73].

PBXGen can build two classes of microstructures: 2D microstructures with columnar grains and 3D microstructures. While the latter is more realistic, the former enables larger feature sizes for a given computational budget. Regardless of the microstructure type, the inputs include material selection (HE and binder types), relative amounts for each (by weight percent), polymer chain molecular weight, HE crystal shape and size distribution, and desired simulation cell dimensions. With this information, the building process proceeds along these steps: 1) Cutting HE particles from their single crystal structure with the desired shape and size distribution; 2) Coating HE particles with polymer chains following the desired length distribution and weight percentage; 3) Packing coarse-grain particles representing coated grains following the predetermined distribution; 4) Replacing coarse-grain particles with all-atom coated particles followed by a relaxation and compaction of the all-atom PBX to the desired density under the desired thermodynamic conditions. Figure 1 illustrates these key steps to generate an all-atom PBX system.

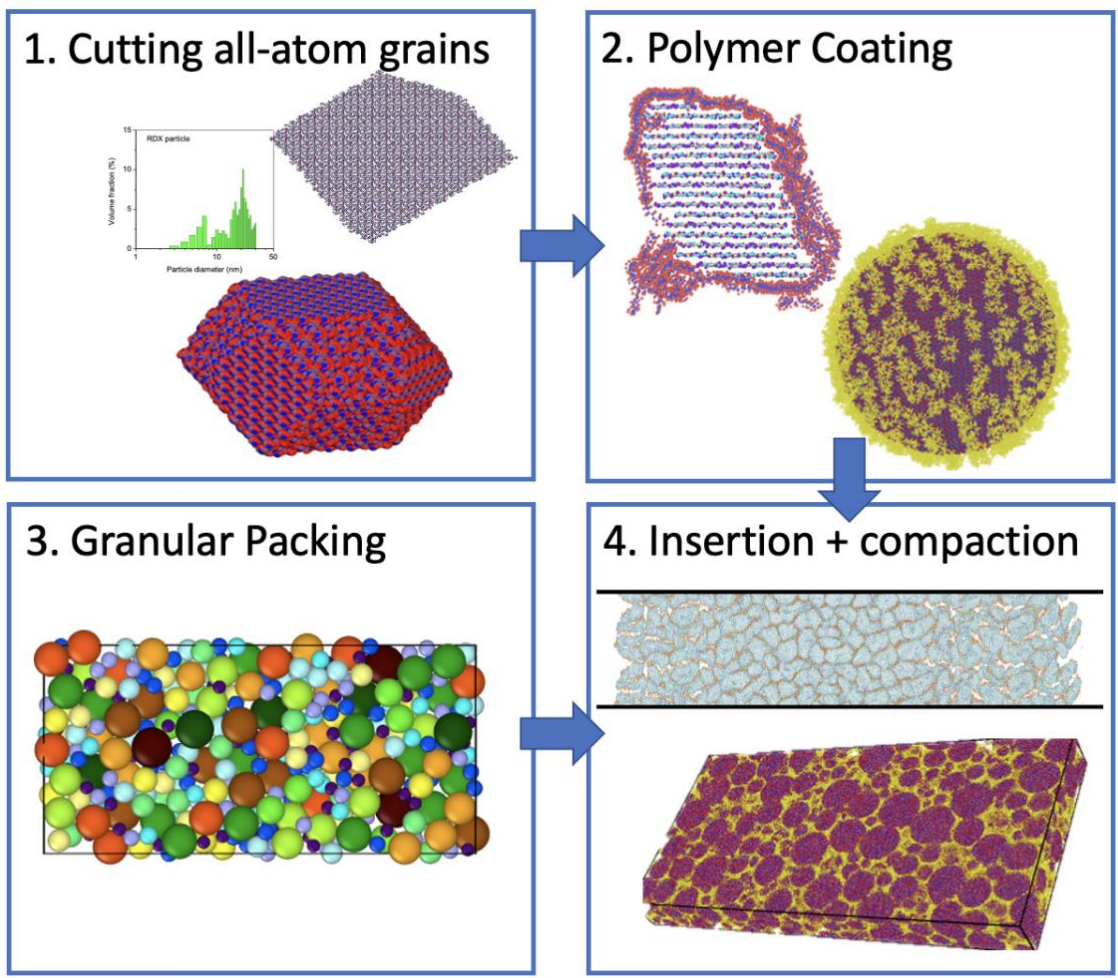

*Figure 1: Flow chart of PBX builder: PBXGen*

### 3.1 Step 1: HE crystalline particles

In this step, individual HE particles are created by cutting the desired geometries and sizes from periodic single crystals. Several approaches are utilized to attain realistic grain morphologies. The first approach is to use either simple, geometric shapes, such as cylinders (columnar microstructures) and spheres (in 3D), or to cut a faceted particle with various surfaces that match experimentally grown crystals [74,75]. Crystal morphology depends on growth conditions and processing, e.g. milling can significantly affect morphology. During crystal growth, morphology is mainly determined by the relative growth speed between the various surfaces. Growth speed is often dominated by the attachment energy (negative in value), i.e. the energy associated with adding a molecule to a flat surface [76]. The more negative the attachment energy, the faster the corresponding surface will and the smaller area they will take in the final structure. Thus, the morphological importance (MI) of a crystal face can be estimated by assuming its growth rate is proportional to the absolute value of attachment energy. An estimate of the relative MI (the maximum relative MI is 1) for each observable face is determined based on attachment energy, which we obtain from MD simulations in Ref. [70]. Given selected particle diameter from the size distribution, the radii of each face are then set to be proportional to the relative MI.

However, grain morphologies are stochastic in nature, with several other factors affecting the crystal growth. To consider this stochasticity, a second approach is to use Monte Carlo (MC) simulations based on the Metropolis algorithm [77] to determine a variety of potential grain shapes. Initially, a set of available free surfaces is predefined with the surface energy listed for each one, where surface energy is related to milling and surface cleavage. In the first MC step, a preliminary particle is created using randomly chosen surfaces from a predefined list with constraints that creates a closed polygon. The total energy of the particle is calculated as the sum of the energy contributions from each of the free surface and the cohesive energy contribution from the volume. In the following MC steps, two moves are proposed. One that alters the radius of one of the free surfaces to the center of the grain and another that potentially alters which crystallographic free surface it is, choosing from a geometrically neighboring surface in the set of available surfaces. The radius is altered based on a normal distribution and the surface choice a uniform distribution, making both steps reversible. After the move is randomly selected, the new configuration is accepted or rejects following the Metropolis algorithm; this generates grains following the Maxwell-Boltzmann distribution. Figure 2 shows examples of grain morphologies, with the rightmost stemming from the MC approach.

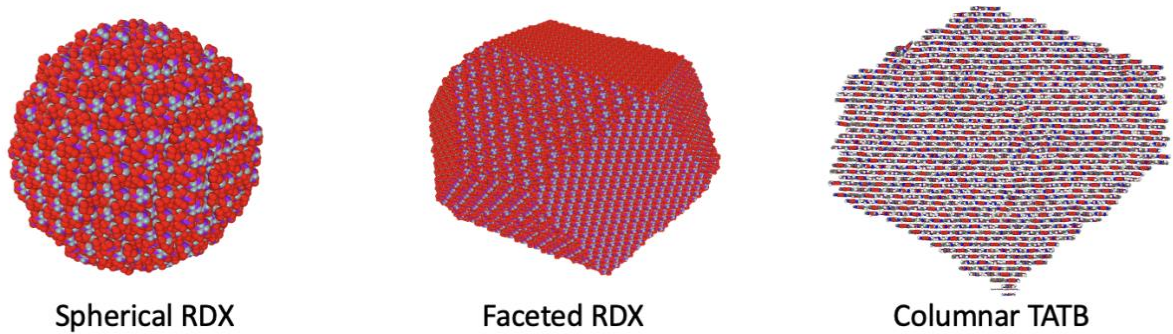

*Figure 2: Various HE particles cut from a set geometries (left two) and a Monte Carlo Approach (right).*

### 3.2 Step 2: Coating HE particles with polymers

Experimentally, HE and polymer binders are typically mixed via a slurry process in which polymer powders and HE powders mixed in an organic solvent [73]. However, on MD timescales, the process of polymers wetting the surface of HE grains must be artificially accelerated. Given a HE particle, the number of polymer chains needed to give the correct weight percentage for the PBX is calculated. For the columnar microstructure, it is computationally easier to generate polymer chains directly in the vacuum surrounding the HE particle by using simple Configurational Bias Monte Carlo approach [78].

For the 3D microstructure, where significantly more polymer is needed, we start by placing coarse-grained (CG) beads, each representing a globular polymer chain, on the surface of a larger CG particle representing the HE grain. The polymer CG beads are placed on the CG particle surface one by one, randomly, and avoiding overlap, until the desired number of CG beads is reached. If the predetermined amount could not be reached after a preset maximum number of iterations, the radius of the CG particle for the HE grain is slightly increased. After all CG beads are placed, pre-equilibrated all-atom polymer chains are inserted into the CG bead locations and the all-atom HE grain is placed into the center of the large CG particle.

To accelerate the coating process, inwards velocities are assigned to all polymer atoms with velocities in the direction of the center of HE particle and the initial value in the range of 0.5 km/s to 1.0 km/s. Converging velocities can be assigned multiple times with adiabatic (NVE) dynamics run in between. The coated particle is then equilibrated under NVE condition, with rigid HE atoms and reassigning polymer velocities according to a Maxwell-Boltzmann distribution. The system is then fully thermalized with fully flexible atoms in both materials with dynamics run under isothermal isochoric (NVT) conditions. Figure 3 shows coated, all-atom HE particles. We note that the CG bead procedure works only for short chain lengths were a spherical CG bad can represent it. For higher molecular weight systems in which entanglements and more complex chain geometries exist, the former method of placing chains farther away in the vacuum is required.

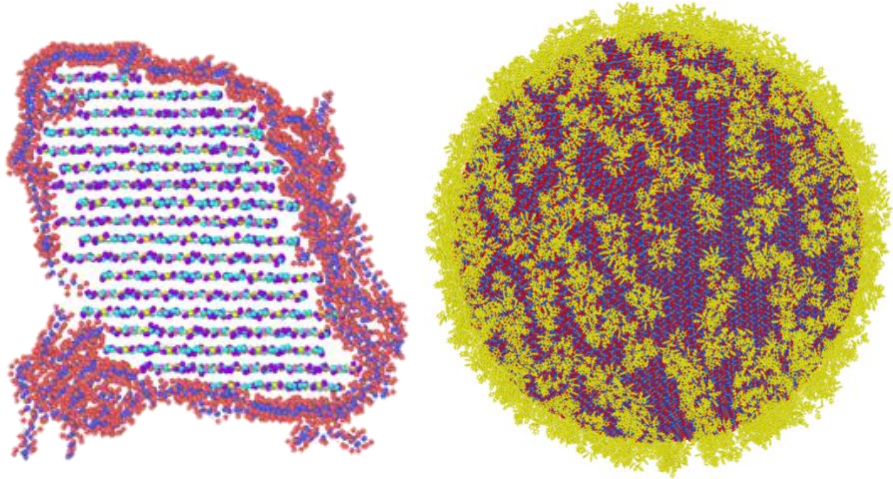

*Figure 3: Polymer-coated HE particles. Left: coated cylindrical TATB grain with 7% PS by weight. Right: coated spherical RDX particle with 10% PS by weight.*

### 3.3 Step 3: Packing CG microstructure

Due to the high computational cost of packing and compressing the totality of the all-atom grains, an initial packing using CG particles is performed, where the particles follow the size distributions of the coated all-atom HE grains. The CG grains are packed into the system cell using the granular dynamics package [79] as implemented in LAMMPS. CG particles with different diameters are randomly poured into the box and the box dimensions can be adjusted to make sure the entirety of the desired grains are placed into the box.

After all CG particles are placed in the simulation cell, granular dynamics under NPT conditions is conducted to get a high packing fraction with the desired box dimensions. Packing fractions can readily reach around 0.65. A bimodal particle size distribution used to create one of the RDX-based PBX systems and the resulting packed CG system are shown in Figure 4.

Note that most PBXs exhibit a bimodal size distribution [22,71] with large grains in the order of 100 microns and small grains in the range 0.5 to 5 microns. These sizes are not achievable with atomistic simulations and the example in Fig. 4 shows a microstructure with a scaled down bimodal distribution affordable for all-atom MD simulations.

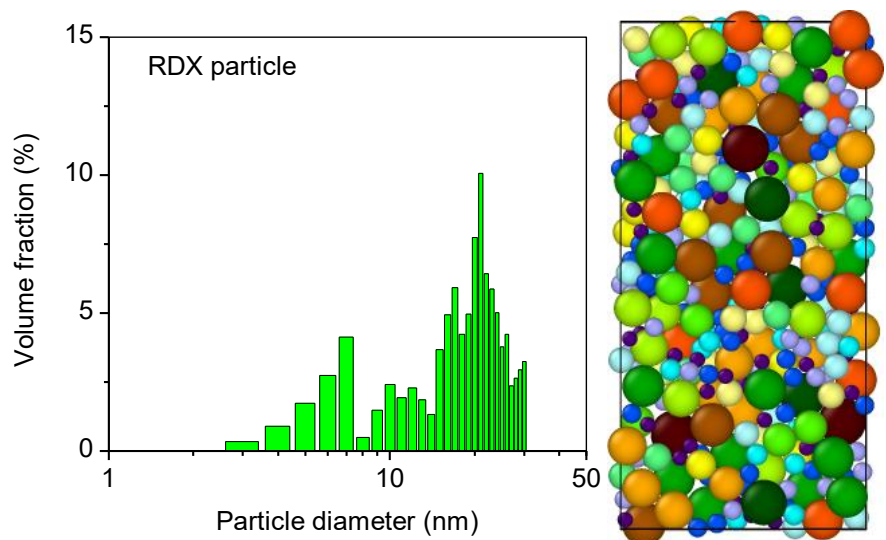
*Figure 4: A packed system with CGs representing coated HE particles which diameters in a bimodal distribution*

### 3.4 Step 4: Packing, equilibration, and compaction

The next step is to place the coated particles into the CG microstructure. Each packed CG particle is replaced with a corresponding coated HE particle, and a resided rotation can be applied, typically either random or no rotation. Figure 5 displays the final packed all-atom systems after CG replacing. It can be seen that there are considerable spaces in between coated HE particles. The overall mass density is typically in the range of 0.3-0.4 g/cm$^3$; therefore, additional compaction is necessary.

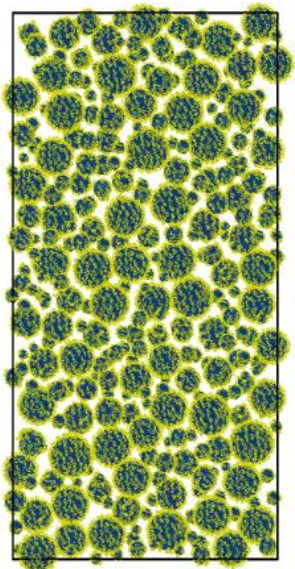
*Figure 5: Packed systems with all-atom coated HE particles*

Compaction is the most computationally expensive part of the building process as it is the only step that runs on the entire all atom system, requiring several hundred picoseconds of run time to perform properly. Initially, the system is thermalized to a high temperature (~600K) under isothermal isochoric conditions (NVT), keeping the HE atoms fixed for ~20ps. Isobaric isothermal (NPT) are then conducted at moderate temperature (~450K) and moderate pressure (~50MPa) bring the system to a compacted density without storing much strain energy within the individual grains. To save computational cost, the relaxation is separated into three stages. In the first stage, the long-range electrostatic interactions, which are very time consuming, are turned off until the density reaches a stable value. In the second stage, the long-range electrostatic interaction is turned on and NPT dynamics continue until equilibrium is again reached. Finally, the NPT dynamics are set to 300K and 1 atmosphere for a final equilibration to ambient conditions. For the RDX/PS system, these three stages take roughly 300ps, 200ps and 50ps, respectively, reaching a density of about 1.5g/cm$^3$.

In some circumstances, such as impacting for shockwaves, a system with a free surface or non-periodic boundary is desirable. To compress a PBX system whilst leaving a surface free involves setting a simulation cell with artificially extended boundaries in one direction, without applying pressure or explicit deformation in that dimension. Other dimensions, with periodic boundaries, are still compressed as described above. For the non-periodic dimension, an inward velocity is applied on top of thermal velocities for grains near the free surfaces. This inward velocity is kept in the subsonic regime (100-200 m/s) to minimize impact damage. Overall, this prevents the transverse pressures from extruding or ejecting particles to alleviate stress. This non-periodic compaction potentially leads to a complex microstructure where the central region of the system reaches a higher density, while the regions close to free surfaces leaves with a higher level of inter-grain porosity, see Figure 6b. This will be taken advantage of in this work to inspect the shock response of different microstructures in a single simulation.

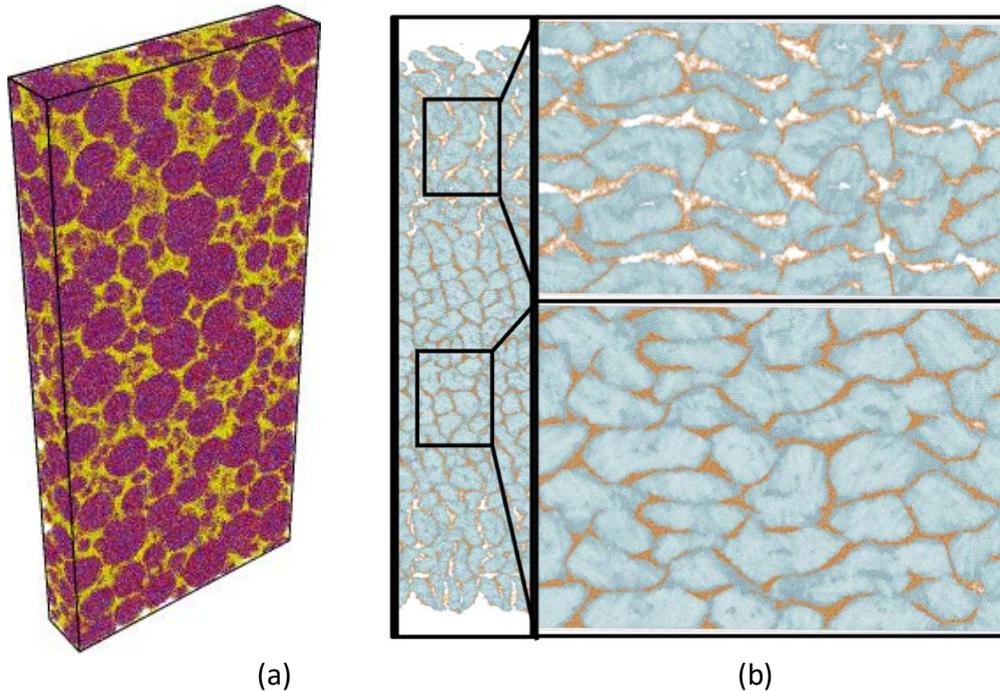

(a)            (b)

*Figure 6 Compacted all-atom PBX systems: (a) RDX/PS 3D system, (b) TATB/PS pseudo-2D system with free surfaces*

# 4. Mechanical properties

PBXGen is utilized here to generate a set of 3D microstructures consisting of 1,3,5-Trinitro-1,3,5-triazinane (RDX) and Polystyrene (PS) with two different RDX/PS weight ratios: 83/17 and 90/10, and a columnar microstructure with 2,4,6-triamino-1,3,5-trinitrobenzene (TATB) and PS with the TATB/PS weight ratio 93/7.

The columnar TATB system involves grain diameters following a normal distribution with an average of 15 nm and a standard deviation of 2 nm. PS chains are all 40 monomers in length. A free surface is left in the shock direction, which is about 350 nm in length, with a width of 60 nm. In the central, well packed region, the density is 1.73 g/cm$^3$, where TMD is 1.93 g/cm$^3$ and experimental samples of PBX 9502 can have densities near 1.89 g.cm$^3$ [80].

The diameter of RDX particles range from 5nm to 22nm and number of particles of each size is determined based on the bimodal distribution shown in Figure 4. The PS chains are all 40 monomer long (molecular weight ~4168 g/mol). Four different RDX/PS systems are created, varying RDX particle shape and orientation, as well as polymer percentage. The cell dimensions are approximately 60nm x 20nm x 100nm with the total number of atoms around 10 million. Table 1 summarizes the various RDX-based PBXs and the resulting mechanical properties.

*Table 1: PBX systems and mechanical properties*

| PBX composition | Density (g/cc) | Porosity (vol%) | Young's modulus (GPa) | Tensile yield stress (MPa) | Tensile yield strain (%) | Shear modulus (GPa) | Shear yield stress (MPa) | Shear yield strain (%) |
|---|---|---|---|---|---|---|---|---|
| RDX/PS: 83/17, spherical, oriented RDX | 1.5085 | 5.65 | 8.1 ±1.0 | 148 ±8 | 5.9 ±0.9 | 1.9 ±0.1 | 87 ±5 | 8.4 ±0.6 |
| RDX/PS: 90/10, spherical, oriented RDX | 1.4992 | 7.14 | 8.8 ±0.9 | 178 ±8 | 6.6 ±1.0 | 2.1 ±0.1 | 92 ±6 | 7.7 ±0.7 |
| RDX/PS: 90/10, spherical, random RDX | 1.4926 | 8.06 | 7.3 ±0.7 | 157 ±7 | 6.5 ±1.1 | 3.0 ±0.1 | 92 ±4 | 7.5 ±0.6 |
| RDX/PS: 90/10, faceted, random RDX | 1.4909 | 8.07 | 7.4 ±0.8 | 160 ±5 | 6.6 ±1.0 | 2.6 ±0.1 | 92 ±6 | 6.9 ±0.5 |

The final densities of these systems are all approximately 1.5 g/cm$^3$. The ideal, void-free density for RDX/PS systems with a weight ratio 90/10 is 1.72 g/cm$^3$. However, the maximum density typically reached experimentally is about 1.58 g/cm$^3$ [76]. The smaller density in our systems as compared to experiments is expected due to the smaller grain size. The porosity for these systems ranges from 5.5% to 8%, larger than the experimental values, in the range of 2-3% [67]. We note that increasing the polymer binder fraction reduces porosity as more space between HE grains can be filled.

To quantify the mechanical properties of our microstructures, uniaxial and shear deformations were conducted at 300 K with a strain rate of $2\times10^9$ s$^{-1}$. Figures 7 and 8 show the stress-strain curves for uniaxial and shear deformations, respectively. The resulting values of moduli, yield stresses, and yield strains are listed in Table 1. The moduli are obtained by fitting the stress-strain curves at low strains which is taken to be the elastic region. The Young's moduli range from 7.3 to 8.8 GPa, in excellent agreement with the experimental data 8.6 GPa for PBX9407 (RDX 94%, Exon461 6%) [76]. The open symbol curve in Figure 7 is experimental data for PBX9501 (HMX 95%, Estane 2.5%, BDNPA-F 2.5%) under compression [81].

The strong role of microstructure is clear in the tensile yield stress, which varies from 148MPa to 178MPa. Comparing the two microstructures with oriented spherical grains, we observe high strength with increasing crystal load. We attribute this to the larger percentage of polymer with a lower strength. Finally, a reduction in strength is observed in microstructures with random crystal orientations. While additional analysis is needed to fully understand this result, we attribute the lower strength to anisotropic elasticity resulting in more heterogeneous stress distributions in the randomly oriented PBX. The resulting stress localization can result localized deformation and yield.

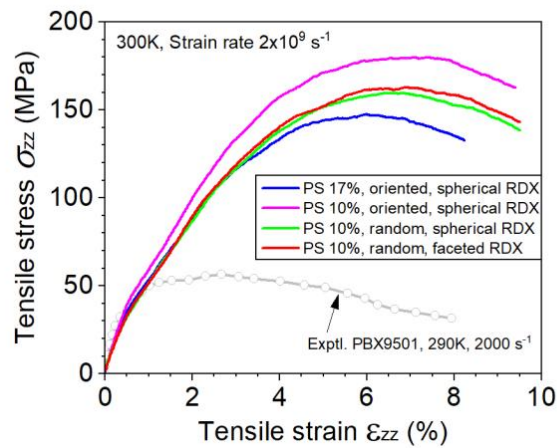

*Figure 7: Tensile stress-strain curves of RDX/PS systems*
*(experimental data is for PBX9501 compression taken from Ref. [81])*

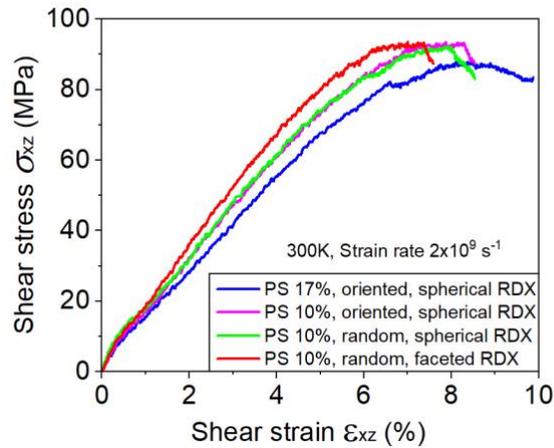
*Figure 8: Shearing stress-strain curves of RDX/PS*

# 5. Shock response

Understanding energy localization into hotspots during shock loading is one of the main open science questions motivating our development of PBXGen. In this section, we discuss our initial results of the response of our PBX systems to shock loading. Section 5.1 uses a computationally efficient simulation technique to characterize the Hugoniot equations of state of our RDX-based PBX systems. Sections 5.2 and 5.3 focus on explicit shock propagation simulations in RDX and TATB systems, respectively.

## 5.1 RDX/PS Hugoniot equation of state

Hugoniot curves are generated using the constant stress, uniaxial, Hugoniotstat method [82]. This technique allows the system to evolve to a desired shocked state by coupling a barostat and thermostat designed around the Hugoniot jump conditions. Shock pressures studied range from 1 to 30 GPa with 20 ps of runtime and a time step of 0.25 fs. Quantities of interest are averaged over the last 10 ps. Figure 9 displays the shock velocity vs. particle velocity, $U_s$-$U_p$, temperature vs. $U_p$, and pressure-$U_p$ relationships. All PBXs studied exhibit similar Hugoniot curves, indicating insensitivity to microstructural details. This is because the Hugoniot depends on the compressive response of the material.

Due to the polymer binder, as expected, the PBX systems have considerably lower shock velocities than that of single crystal RDX (under the same Hugoniot conditions with Smith-Bharadwaj potential [66]). These results are also slightly higher than the experimental data for an RDX based PBX9407 [83]. It follows suit from the $U_s$-Up relations that single crystal RDX has a much higher pressure response. Interestingly, the shock temperature for the PBXs is slightly higher than that of the single crystal RDX due to the inclusion of porosity in the PBX systems.

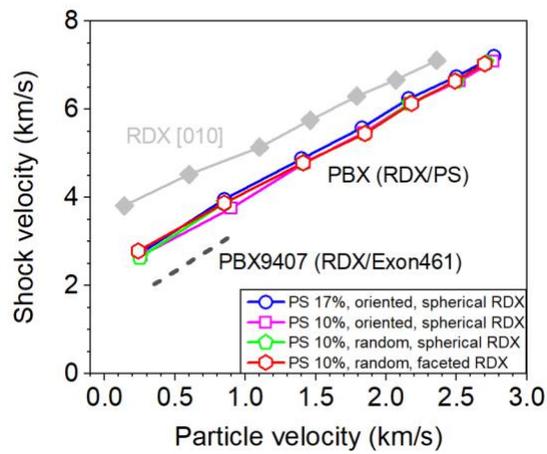

(a)

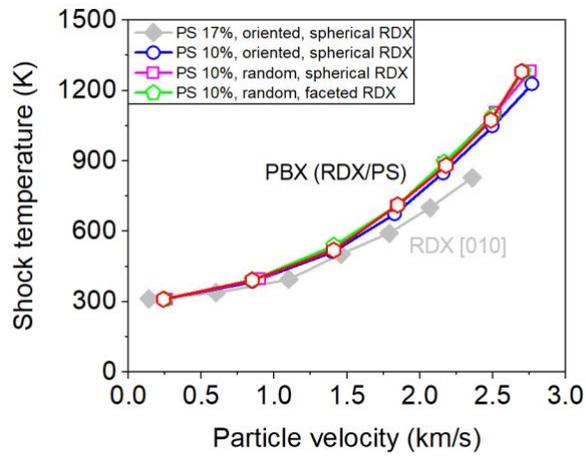

(b)

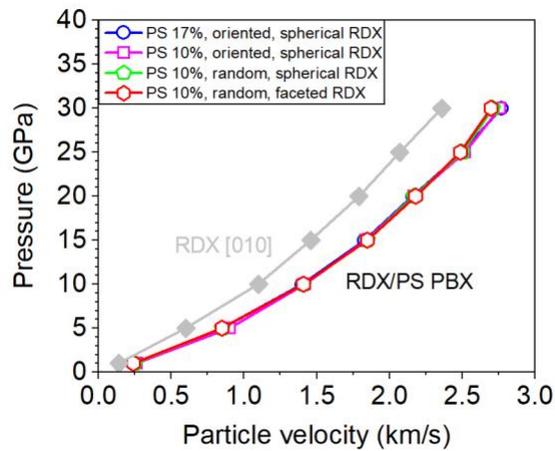

(c)

*Figure 9: Shock Hugoniot for RDX/PS PBXs: (a) Shock velocity; (b) Shock temperature; (c) Shock pressure*

## 5.2 RDX/PS hotspot formation

There are two ways to simulate the propagation of shock waves in molecular simulations. The first, and most common method, is ballistic impact where an explicit collision between a projectile and a target are simulated using adiabatic MD. This requires samples with free surfaces and involves a rigid and infinitely massive piston [84]. An alternative approach which is applicable in 3D periodic systems, is a converging shock method, in which the outer simulation cell boundaries in the shock direction are driven inwards at a velocity of $2u_p$, remapping atomic velocities as atoms cross the moving boundary [85]. To avoid the creation of free surfaces, we used this latter technique to shock our RDX PBX samples. Figure 10 shows a temperature map of one of the RDX/PS systems with 10% PS and oriented spherical RDX particles. The particle velocity for this simulation is 2.5 km/s and the temperature range shown is in the range of 300K to 2500K. Despite the system being replicated once in the shock direction, such that each wave transverses the same microstructure, the waves are passing microstructural defects (mainly porosity) from two different directions, leading to hotspots that are not identical in size or temperature.

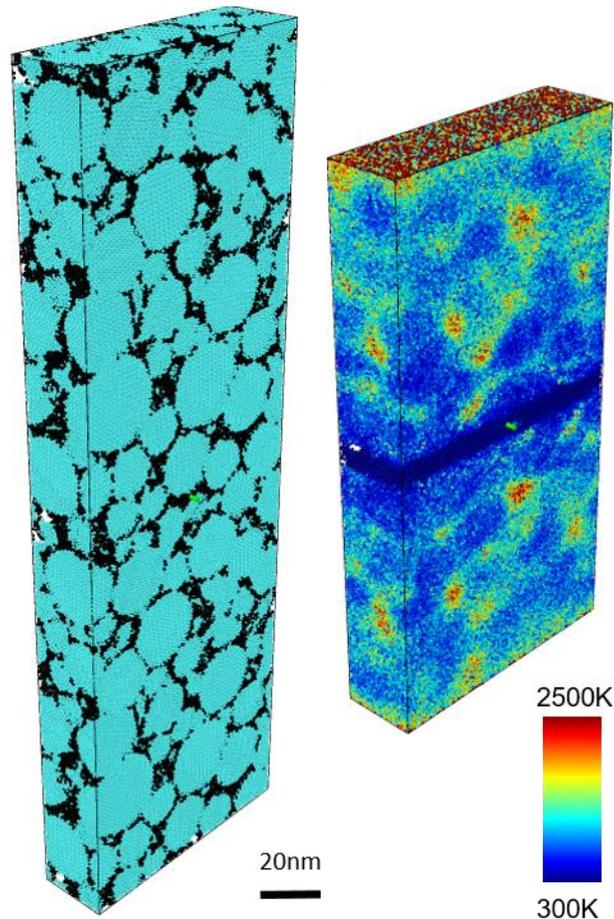

*Figure 10: Hotspot distribution in a RDX/PS system (shock in the vertical direction)*

## 5.3 TATB-PS: shock loading

The columnar TATB/PS system, built with a free surface and a graded microstructure, is impacted utilizing a reverse ballistic setup [84] with an explicit piston (momentum mirror) at the bottom boundary. TATB/PS particle velocity was set at 1.5 km/s. Figure 11 shows molecularly averaged (monomer averaged for PS) maps of particle velocity at various times in the simulation. For early times when the wave is in the in the first porous region (10-30 ps), the wave front is highly disparate across the width of the sample, which is likely to do with the large amount of surface roughness on the impact plane, seen in the leftmost panel of Figure 11. The wavefront does not reach a clean, nearly planar shape until it reaches the closely packed central region of our sample (40-60ps). At this point, a two wave (elastic-plastic) shock structure can be seen, with plastic particle velocities mostly independent of the grain despite high levels of anisotropy in TATB [86]. In the final two frames, when the shock front re-enters a porous region, molecular jetting into voids accelerates the wave locally, which leads to the bands of higher velocity (red) regions seen in the 80ps frame.

Figure 12 shows 1D wave profiles of temperature and density for the same 1.5 km/s shock. As expected, the temperature profiles in the outer sections of the plot (regions with higher porosity) are considerably higher than in the center (packed region) due to the formation of hotspots.

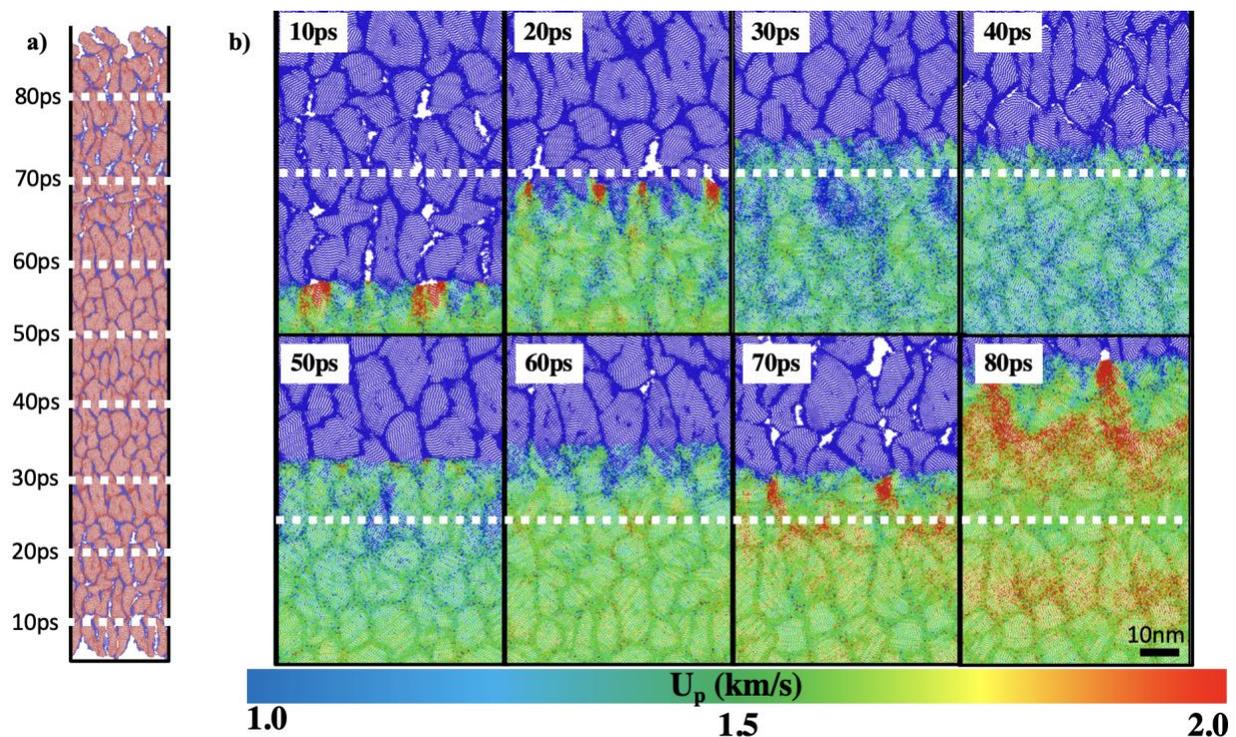

*Figure 11: Molecular particle velocity maps for various times for the TATB+PS PBX for a shock of 1.5 km/s. White line across panel b frames correspond to locations of the white dashed lines in the unshocked full system of panel a.*

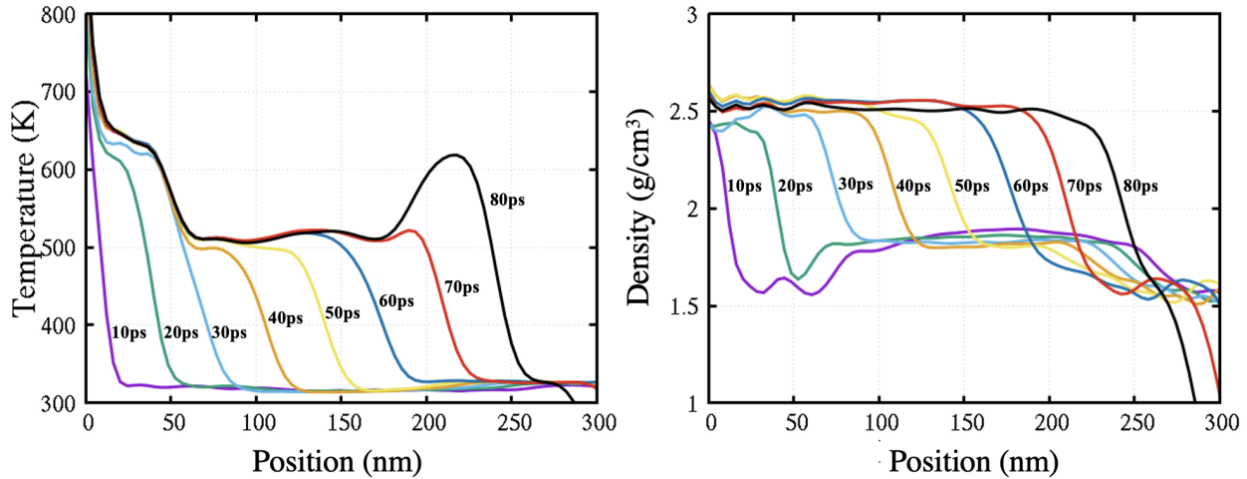

*Figure 12: 1D temperature and density profiles for the 1.5 km/s shock in the TATB+PS PBX which each curve representing 10 ps intervals from 10-80 ps.*

Figure 13 shows a composite mapping of all molecular/monomer temperatures for the columnar TATB PBX. This is done by taking the position and internal (rotational-vibrational) temperature at every molecule and monomer at 5ps after being shocked, where shock time is defined as the first time the molecule/monomer has a velocity in the shock direction greater than 1.0 km/s where $U_p$ for the shock is 1.5 km/s. This shows the clear and distinct increase in temperature in the porous regions, marked as outside of the dashed red lines. These hotspots are entirely from inter-granular pore collapse as no defects are engineered within the grains during the building process. By comparing the packed and porous regions there is a distinct difference between the temperatures due to hotspots in the porous as well as ejecta re-shocking grains at higher velocity.

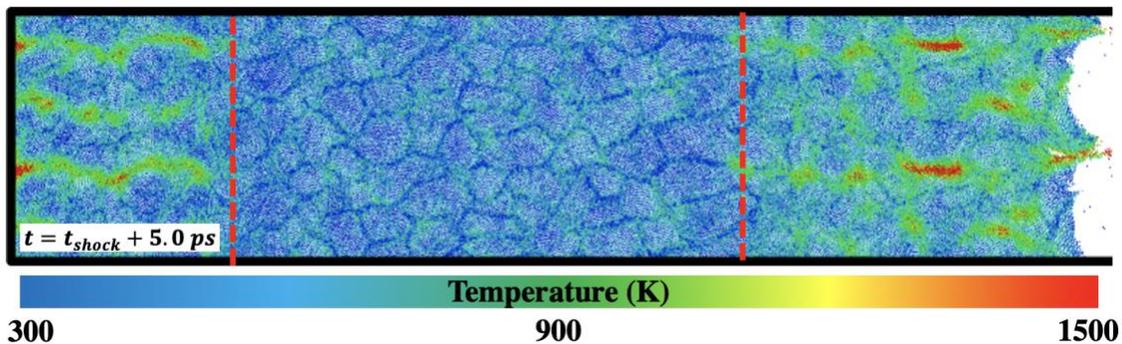

*Figure 13: Composite dump for the TATB+PS PBX with each molecule/monomer rendered at its center of mass and colored by rotation-vibration temperature for 5ps after each molecule/monomer experiences the shock*

Figure 14 shows distributions of the temperatures of TATB molecules in the two regions at 5ps after shock. Not only is the tail of the porous region much longer, but the most probable temperature is roughly 100 K higher in the porous region. On the 5ps timescale, thermal transport will only play a minor role in this. Most likely, the shock speed / particle velocity in these grains is increased due to the ejecta into the pores re-shocking downstream, leading to higher local temperatures in the grains as well as in the hotspots.

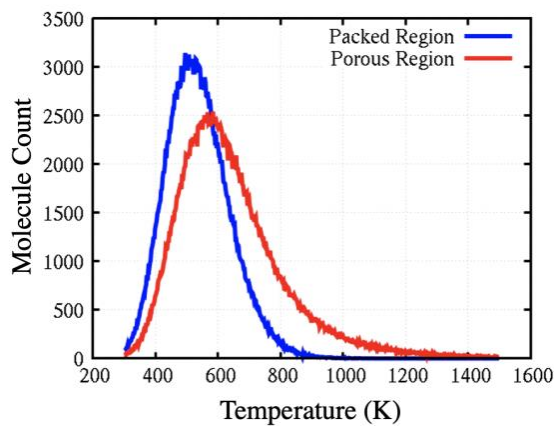

*Figure 14: Distribution of HE molecular temperatures in the TATB+PS PBX for times at 5ps after each molecule is shocked*

## 6. Conclusions

We introduced PBXGen, a generally applicable all-atom builder for PBX microstructures for molecular dynamics. Our method starts from the desired size distribution of the HE grains and cuts from single crystals. These crystals are then coated with polymer of the desired molecular weight, with the amount of polymer determined by mass ratio. Given the size of the coated particles, we use granular simulations to pack grains at a coarse grain representation. The final step is to insert the all-atom polymer-coated HE grains into their corresponding CG grain and system is equilibrated and compacted. We demonstrated PBXGen by building a columnar PBX system based on TATB and multiple 3D RDX based microstructures with faceted and spherical particles. We characterized mechanical properties of the RDX systems via uniaxial deformation simulations and compared the response against experimental stress-strain curves. We find that microstructure has a strong effect on ultimate tensile strength of the PBXs, and as expected, a weaker effect on elastic properties. We also performed shock simulations on both the TATB and RDX systems. The shock velocity vs. particle velocity relationship in the RDX system is in good agreement with experiments. Both the TATB and RDX systems are shocked at high impact velocities to induce significant hotspots in which the resulting thermal localizations are correlated to the initial microstructures, with inter-granular porosity playing a big role in the development of large hotspots.

## Acknowledgments

This work was supported by the U.S. Office of Naval Research, Multidisciplinary University Research Initiatives (MURI) Program under Contract No. N00014-16-1-2557, program managers Chad Stoltz and Kenny Lipkowitz.